# SECURITY CONCERNS OF IPv6-SLAAC AND SECURITY POLICIES TO DIMINISH THE RISK ASSOCIATED WITH THEM

*Parul Sharma*
*Research Scholar, Sedulity Solutions & Technologies, New Delhi*
*parul_sharma@outlook.com*

*Dr. Anup Girdhar*
*CEO-Founder, Sedulity Solutions & Technologies, New Delhi*
*anup@sedulitygroups.com*

**ABSTRACT**

*Though, Ipv6 has presented a long list of features that will actually change the entire networking environment, it still represents a very small proportion of Internet traffic. One of the major reasons behind its partial acceptance is its security, as these features were incorporated in Ipv6 just to enhance its overall performance and quality of service, thereby neglecting its security part. One such feature is* **Stateless Address Auto Configuration** *or simply* **SLAAC,** *which have undoubtedly reduced a lot of overhead in configuring networks, but on the same time it has challenged the security of Ipv6 networks. This research work, therefore targets at proposing some security policies to mitigate the risks associated with SLAAC, by understanding its working scenario and trying to create the possible attack vector to highlight its security concerns. Hence, the implementation of these policies will directly add security to networks and make them resistant to SLAAC based attacks.*

## I. INTRODUCTION

Using IPv6 [1], [2], a host machine can generate its own IP address, without any need for configuration or need of any additional address configuring servers. This automated process of address generation is termed as Stateless Address Auto Configuration or simply SLAAC [1], [2], [3].

SLAAC allows host devices within IPv6 networks to configure its own IP address by combining locally available information and parameters found in Router Advertisements or simply RA's [4]. IPv6 enabled routers advertise information that identify the network associated with it, while hosts generates a unique interface identifier. A combination of both forms an IPv6 address. SLAAC therefore has provided a massive accountability to routers, which in point of fact has given birth to SLAAC-based attacks.

Such attacks can involve an attacker to kill the default router to introduce a DOS [5] factor, or make a fake router to become the default router of the network to fetch the entire network traffic and introduce MITM [6] factor. Further, more complicated SLAAC based attacks can involve an attacker to forward the fetched traffic to a fake DNS [7]. So once the host devices get auto-configured for IP addresses using these fake router advertisements, the attacker can directly control the network as per his malicious intentions.

In this paper, I therefore present the possible attack vectors to figure out vulnerabilities in SLAAC. By its end, my research work concludes by stressing on some security policies, which when implemented will make networks much more secure and resistant to such SLAAC based attacks.

The rest of this research work is organized into following sections: Section-2 gives an overview of the related work; Section-3 focuses on the working of SLAAC; Section-4 highlights the security concerns of SLAAC and possible attacks; Section-5 presents the proposed security policies and concludes the research work.

## II. RELATED WORK:

The related work to Ipv6-SLAAC can be categorized into following 3 categories: a) Basics of IPv6 and its features; b) IPv6 vulnerabilities and its security; c) IPv6 addressing issues. [20] Microsoft TechNet, in 2008, explored IPv6 and its features including address auto-configuration. [21] Jinesh Doshi, Rachid Chaoua, Saurabh Kumar and Sahana Mallya did a comparative study of IPv4 and Ipv6 to highlight the co-existence in both the technologies and the adaptation strategies related to it. [22] Shahnawaz Sarwar and Aiman Zubair focused on IPv6 addressing and IPv6 routing protocols. [23] Harith Dawood focused on the security vulnerabilities of IPv6. [24] Clinton Carpene and Andrew Woodward exposed the IPv6-addressing concerns and privacy risks associated with them.

Finally I found that the research work has been done on IPv6-feature suite and its vulnerabilities on a very general ground. Individually, much research is not done on SLAAC and its security implications. This research, in particular, focus at security concerns of SLAAC and propose policies to develop a mitigation system against them.

## III. SLAAC:

Ipv6 is popular for its numerous benefits to end users. Gone are the days of Ipv4, when end users needed to configure their machines manually or at least configure their DHCP [8] server to reduce a bit of overhead. Ipv6 has introduced the concept of 'hot plugging', which will allow a host to configure its Ipv6 address automatically once it is connected to a hotspot within a network.

This feature is technically known as **STATELESS ADDRESS AUTO CONFIGURATION,** and will automatically configure host machines with a unique IPv6 address, even in the absence of DHCPv6 [9].

The process of this address auto configuration can be divided into following two phases:

**PHASE 1**: Generation of Link- Local Address [10].

- The host devices initially generate a link local address, using which they can communicate only within the local network. This link local address is generated by combining the interface address and link local prefix.

- The moment, a link local address is generated, it get verified for its uniqueness on the network before it is assigned to a host device and used for communication. This verification process is carried out by the host by sending Neighbor Solicitation messages containing its link local address. If this address is used by another host device on the network, it will send back a Neighbor Advertisement for address conflict.

- In case there is an address conflict on the network, the process of auto configuration stops. Now this host can be manually configured, or can be supplied with another interface identifier to continue with address auto configuration.

- Once a host device confirms for the uniqueness of its link local address on the network, it can use this address for IP level connectivity within the local network only.

**PHASE 2:** Obtaining Router Advertisements and generation of Global Address [10].

- The Ipv6 enabled routers send Router Advertisements after every small interval of time so that the host device could carry out address auto configuration without the need to contact routers directly. These RA's play a vital role in generating a global address for a host device that can be used globally out of the local network. These router advertisements are actually accompanied by a series of flags that triggers the address auto configuration in real sense. Further it can contain other parameters like sub-net prefix, lifetime values etc.

- The Autonomous-Address Configuration Flag [11] provides stateless address auto configuration and allows the host device to configure its own IPv6 address. As soon as the host device receives an RA with Autonomous-Address Configuration Flag, it triggers the generation of global Ipv6 address.

- The Ipv6 Global Address is at last generated by combining the unique interface identifier and the network/sub-net prefix provided by RA's.

- Now once the global address is generated, it is again checked for its uniqueness. Duplicate Address Detection or simply DAD [12], is a protocol that solves the purpose of checking uniqueness of a global address assigned to a host device.

Once DAD flags an Ipv6 Global Address with a green signal, the host devices associated with this address becomes a part of the IP New Generation Protocol.

## IV. THE ATTACK VECTORS:

Though SLAAC has automated a part of network administration, a gap in its adaptation has left the end users prone to attacks. Thus, this section, in particularly examines this gap to create possible attack vectors and figure out the vulnerabilities in SLAAC.

There is a possibility that an attacker can intercept the communication link between the host machines and the default router, and then misuse the total automation factor in SLAAC. This malicious intention is given a real-time effect in terms of following steps:

**STEP-1:**

The attacker, at first tries to intercept the actual link between the default router and the host machines by altering the Router Advertisements of the default router. The goal behind altering these RA's is to change the router lifetime which defines the time limit of concerned router to act as default router for that network. But before an attacker can alter these RA's, he needs to capture these RA's by becoming the part of the network.

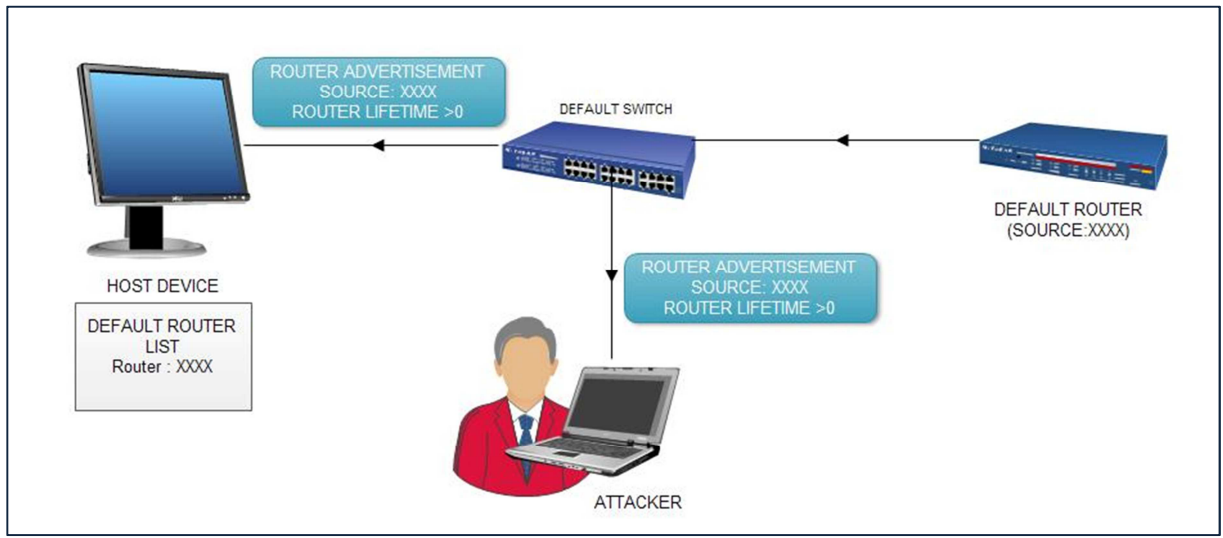

Figure 1: Attacker capturing a valid RA from the default router.

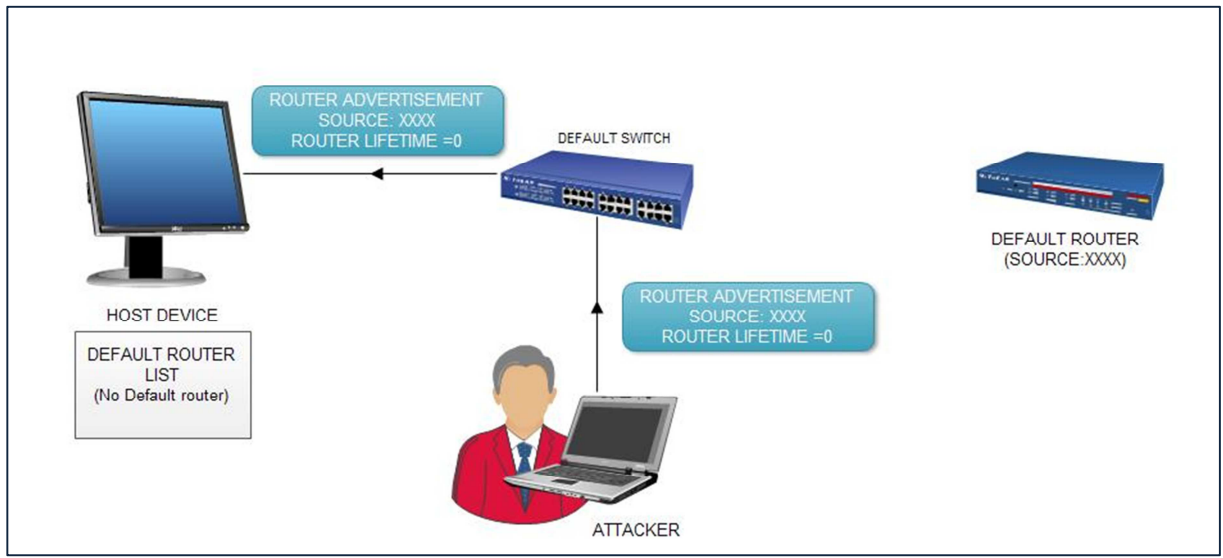

Figure 2: Attacker sending a spoofed RA to host machine to remove its default router.

**STEP-2:**

Now once the attacker is successful in capturing valid RA's from a default router, by changing the router lifetime available these RA's to a value of zero, an attacker is easily able to make the host devices to remove the default router from its routing table as shown in Figure-2.

**STEP-3:**

After an attacker has spoofed the valid RA's and removed the default router from the routing table, it is the right time for him to make his final move. Now at this stage an attacker can proceed with one of the following as per his target and intentions:

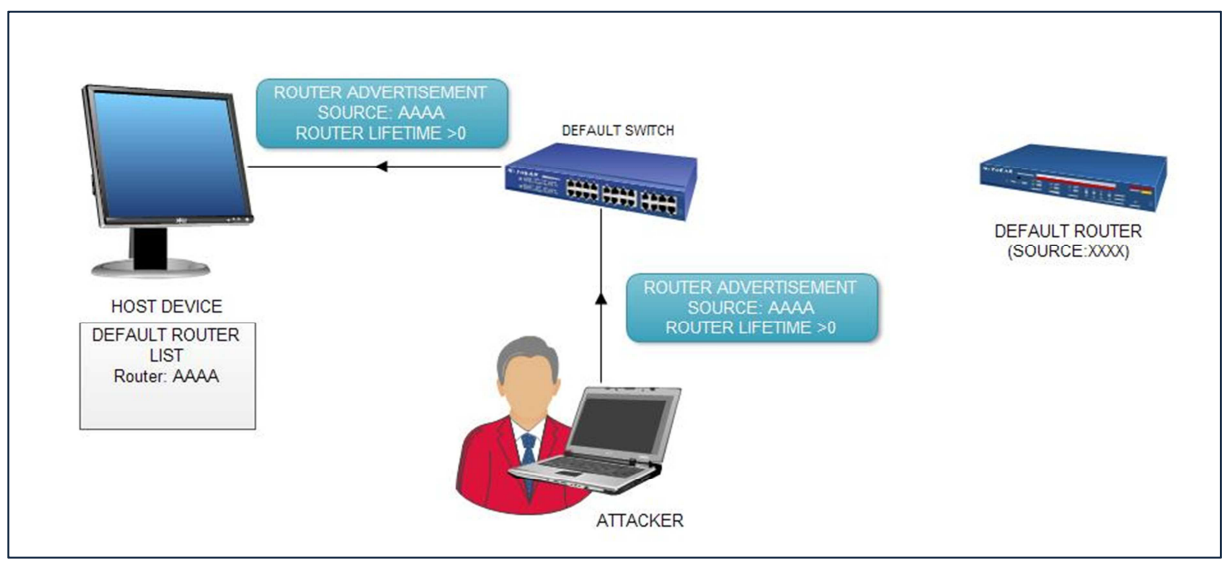

Figure 3: Attacker sending a fake RA to host machine with changed source address.

- **SLAAC- BASED DENIAL OF SERVICE VECTOR:**

Once the attacker is successful to remove the default router from the network, the host machines will have no default route to transfer their packets. It itself is enough to mount a DOS attack. Further, an attacker can also advertise the address of a non-local router or a machine without routing capabilities using fake RA's, which will be installed as the default gateway for host machines once they get configured using SLAAC, as shown in Figure-3. Since these devices will not be able to route the packets to their desired destinations, as they do not exist locally or do not have routing capability, every request by the host device will end up at 'Denial of Services' response by the service provider.

- **SLAAC- BASED MAN IN THE MIDDLE VECTOR:**

An attacker can also introduce a fake router into the network and make the host machines to get auto-configured for addresses using RA's sent by this fake router. Since the network uses SLAAC for address configuration, the total automation factor involved in this process will not allow the end user to know about the transition in the route. Once the host devices get configured using fake RA's, the attacker, being 'man in the middle' is able to fetch the entire traffic to his router and can hamper the privacy of data.

- **SLAAC- BASED DUAL-STACK FAKE ROUTING VECTOR:**

This attack actually targets a fully working Ipv4 network with IPv4 based router to route the packets between the hosts and Internet, as shown in Figure-4. Since most of the operating systems nowadays have Ipv6 enabled by default, they are prone to get configured by an IPv6 enabled fake router using SLAAC.

Hence, this attack vector can target IPv4 networks with IPv6- enabled host devices, and will involve an attacker to deploy a Dual-Stack [13] router with Ipv6 addressing interface on the host side to fetch the entire traffic.

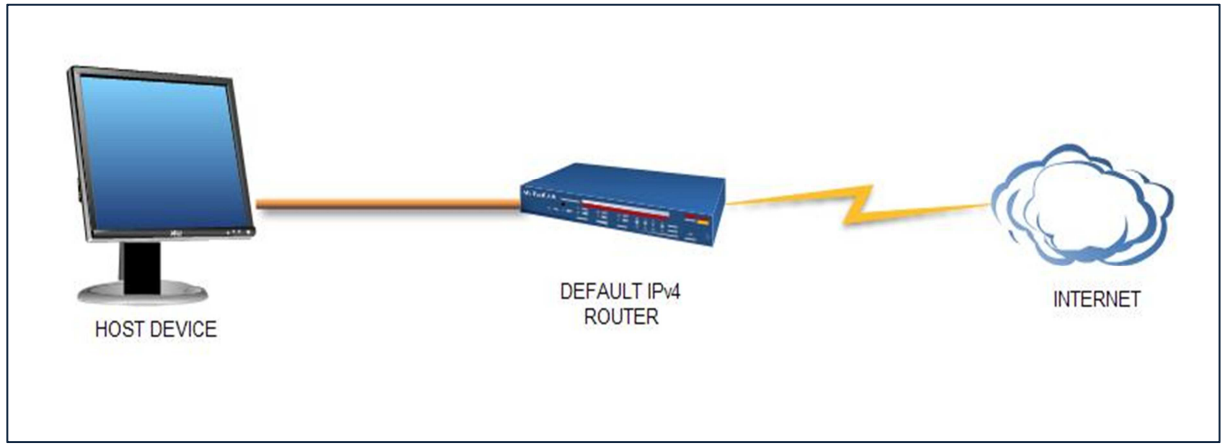

**Figure 4: Vulnerable IPv4 network, with IPv6 enabled host device.**

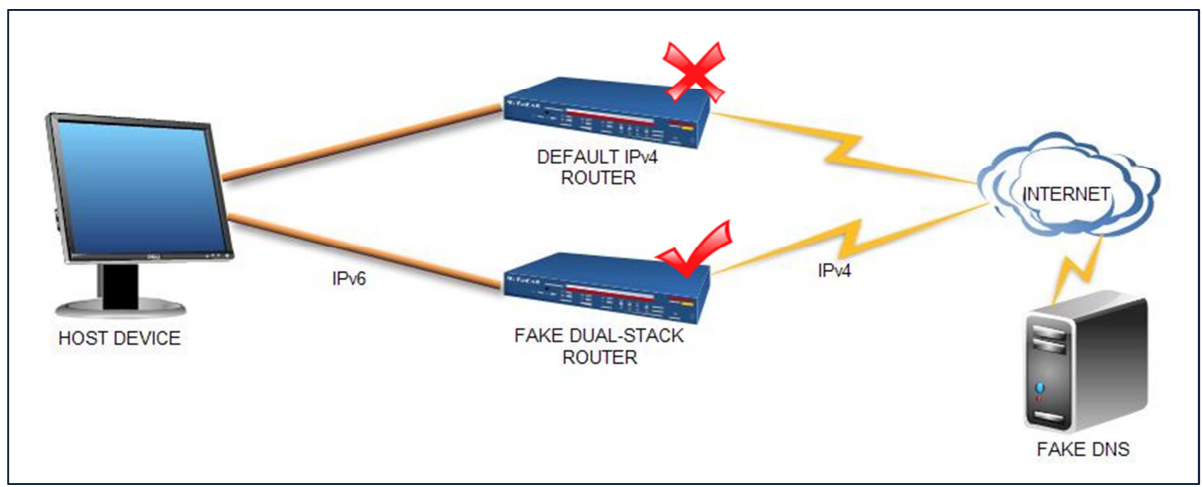

**Figure 5: Deployment of fake Dual-Stack Router to kill default router.**

Since, by default, Ipv6 addressing has a higher priority over Ipv4 addressing, the dual stack router deployed by the attacker will configure the host machines using SLAAC and will become the default router for the network, as shown in Figure-5. Now when an attacker is successful in fetching the network traffic to the router deployed by him, he can easily route this traffic to any spoofed or fake DNS again controlled by him using the concept of NAT-PT [14] on the fake router.

The above cases directly reflect the attacking strategies that can be used to exploit the vulnerabilities in SLAAC. These attack vectors can prove to be highly risky for both IPv4 as well as Ipv6 based networks, as SLAAC itself do not provide any monitoring or mitigating system against such attacks

## V. SECURITY CONCERNS OF SLAAC:

As noted in the previous sections, SLAAC largely depends on routers in the process of address auto configuration. Since routers are entirely the care takers of address assignments under Ipv6 SLAAC, and perform IP based authentication by itself, an attacker can easily spoof a trusted router or even place a fake router to fetch the entire network traffic and employ any of the attack vectors discussed before. The following point's highlights the concerns associated with this huge accountability provided to routers in SLAAC and how these concerns can be used by an attacker to target the security of networks.

- Using the SLAAC mechanism, routers initiate the process of address auto configuration by sending routing information as a part of RA's. These RA's are directly received and processed by host machines without any checks to get an IPv6 addresses. But fake or bogus routing information sent through RA's can substantiate an attacker to request a connection to host machines and capture their traffic.

- Since SLAAC involves RA's to communicate the subnet prefix/ subnet masks to host machines, a massive DOS attack can be launched by communicating the fake subnet masks to host machines. Since a host machine will have to reply for every received RA, no matter whether it was requested or not, a chain of such fake RA's can easily lead to a network crash.

- Under SLAAC, interface identifiers remains unchanged regardless of the network they are associated with. It is so because SLAAC, by default, uses the MAC address as the interface identifier [15]. This directly exposes the host machines to the attacker if he can fetch the entire traffic of the network to a router under his control. This might likewise imply that the host machines are vulnerable against device tracking and additionally simple targets for attacks from anyplace as far and wide as possible. Hence SLAAC, in other words compromise device privacy on the network.

The above concerns clearly highlight the fact that the use of SLAAC can possibly lead to service delays, connection drops and eavesdropping.

## VI. SECURITY POLICIES:

As clear from above section, SLAAC can be a huge risk once the host machines get configured by a malicious router and this can directly target privacy and security of data on the network. Hence it is very important to cover the gap in adaptation of SLAAC to ensure security of networks. The following security policies can play a vital role in diminishing the risks associated with SLAAC based attacks, thereby making SLAAC a much more powerful and secure feature of IP Next Generation Protocol. Such policies will help to identify malicious Router Advertisements [16], as well as will rapidly recover from a state where host machines have already configured using such a fake RA.

- **DISABLE IPv6 ADDRESSING SCHEME IN IPv4 NETWORKS:**

Most of the operating systems today support Ipv6 addressing scheme and have Ipv6 enabled by default. This can invite attackers to take advantage of such vulnerable host devices through the dual-stack fake routing attack discussed in Section-4.

It is recommended to disable IPv6 addressing, since it has no role to play in IPv4 networks.

- **BLOCK FAKE ROUTER ADVERTISEMENTS:**

If router advertisements initiated from unknown sources can be blocked, killing or replacing the default router in the network will become almost impossible, and hence the network can easily be secured from SLAAC based attacks. Usually the first hop for host devices on a network is a switch; therefore by implementing the right set of security features on layer-2 itself, we can block fake RA's. Many good tools are available that work on Layer 2 managed switches. RA Guard [17] is one such tool that monitors Router Advertisements and block unknown or fake RA's. Another way to block fake RA's is to introduce the concept of Access Control Lists once default routers of the network are defined. Using ACL's, only known RA's are allowed to user ports and others can be dropped if they are known to be originated from fake or unknown routers.

- **USE OF SLAAC IN SYNCHRONIZATION WITH SE-ND:**

Under Ipv6, host machines uses Neighbor Discovery Protocol [18] (NDP) for various purposes such as to discover neighboring host machines and routers. Secure Neighbor Discovery or simply Se-ND [18] is the advanced version of NDP that has patched the existing security related flaws in it. Se-ND is a powerful protocol that uses various features to ensure that security of a network is not compromised in anyway. Certification Paths and Cryptographically Generated Addresses are two main features of Se-ND, that ensure authenticity of devices, both the hosts as well as the routers, on the network. So a network architecture that uses SLAAC synchronized with Se-ND protocol can possibly ensure that host machines get configured for addresses using known native routers. Further this practice can even restrict untrusted host devices to connect to the network and cause malicious functions

- **LEGITIMATE ROUTERS SHOULD BE TAGGED WITH HIGH PREFERENCE:**

Router preference allows routers to be classified into three levels: Low, Medium and High. If known and trusted routers can be tagged with high preference, this could suffice to alleviate fake Router Advertisements that were transmitted. For example, Windows Internet Connection Sharing uses Router Preference option to select the router, so if legitimate routers are given high-level of preference, it will be their RA's that will be utilized to proceed for SLAAC. So Router Preference policy might be utilized to recognize legitimate routers and hence can act as a mitigation system against the fake routers.

- **IMPLEMENT '2-HOUR RULE' [19] TO RECOVER AFTER PROCESSING A FAKE RA:**

Once a host device processes a fake RA, it may reach to a conflicting state because of configuring multiple gateways or global addresses. This is the right time when the network administrators should get active and implement '2-hour Rule' proposed in RFC 4862- Section 5.3.3. This policy will help to recover from this critical stage.

## VII. CONCLUSION:

Besides its numerous useful features, organizations and individual users, somewhere lack the trust to use IPv6. The major reason behind this hesitation is the security of IPv6, which is not yet matured enough to shield networks from attacks. Hence there is a need to scrutinize the individual features of IPv6, in order to understand the risks associated with them and develop mitigating systems accordingly. This research work was targeted to explore SLAAC in order to find out the existing vulnerabilities in it and frame out certain security policies that can actually secure the networks that use the concept of address auto-configuration. The proposed security policies will diminish the risks associated with the Denial of Service, Man in the Middle and Dual-Stack Rouge Routing variants of SLAAC- based attacks and hence directly enhance the security of networks.

## ACKNOWLEDGEMENTS

I would like to acknowledge the enthusiastic guidance of Dr. Anup Girdhar, CEO-Founder, Sedulity Solutions & Technologies, throughout the tenure of this research work. Further, I express my sincere gratitude to Mr. Mukul Girdhar, Vice-President, Sedulity Solutions & Technologies, for providing me with every possible resource to complete my work.

Without their support, I would have wobbled through my first step into research. Thank you!